\documentclass[10pt,preprint]{elsarticle}

\usepackage{amssymb}

\begin{document}

\begin{frontmatter}

\title{
Nambu representation of an extended Lorenz model with viscous heating 
}

\author[UHH]{R. Blender} 
\ead{richard.blender@zmaw.de}

\author[UHH]{V. Lucarini}
\ead{valerio.lucarini@zmaw.de}

\address[UHH]{Meteorologisches Institut, KlimaCampus, 
Universit\"at Hamburg, Grindelberg 5, 20144 Hamburg}

\sloppy

\begin{abstract}

We consider the Nambu and Hamiltonian representations of  
Rayleigh-B\'{e}nard convection with a nonlinear 
thermal heating effect proportional to the Eckert number (Ec).  
The model we use is an extension of the classical 
Lorenz-63 model with 4 kinematic and 6 thermal degrees of freedom.
The conservative parts of the dynamical equations 
which include all nonlinearities satisfy Liouville's theorem 
and permit a conserved Hamiltonian $H$ for arbitrary Ec.
For $Ec=0$ two independent conserved Casimir functions exist,
one of these is associated with unavailable potential energy
and is also present in the Lorenz-63 truncation.
This Casimir $C$ is used to construct a Nambu representation of the 
conserved part of the dynamical system.
The thermal heating effect can be represented
either by a second canonical Hamiltonian 
or as a gradient (metric) system using the time 
derivative $\dot{C}$ of the Casimir. 
The results demonstrate the impact of viscous heating
in the total energy budget and in the Lorenz energy cycle 
for kinetic and available potential energy.

\end{abstract}

\begin{keyword}
Rayleigh-B\'{e}nard convection, 
Nambu mechanics,
Lorenz equations, 
Viscous heating
\end{keyword}

\end{frontmatter}

\section{Introduction}

The Nambu representation is an extension
of Hamiltonian dynamics with the aim to
formulate dynamical equations satisfying the Liouville
theorem \cite{Nambu:1973}. In state spaces with dimension above two further
conserved quantities besides the Hamiltonian can be used.
Thus Casimir functions can be considered
as additional Hamiltonians. This concept is useful
in dynamical systems when a physical interpretation
of total energy is not available.
Nambu systems incorporate noncanonical Hamiltonian systems
with the conservation laws being Casimirs of the Poisson
tensor.
Algebraic properties are formulated in terms of anti-symmetric and cyclic 
extensions of Poisson brackets. 
The development of Hamiltonian low-order models
with additional conservation laws 
has been also put forward in terms of gyrostat concept 
\cite{Gluhovsky:1999,Gluhovsky:2006}.
The Nambu representations of hydrodynamics
\cite{Nevir:1993} has shown up to be useful in the
design of numerical algorithms in geophysical fluid dynamics
since they deliver a general tool to preserve 
conservation laws \cite{Salmon:2005, Salmon:2007}.
The Casimirs of ideal hydrodynamics have been
explicitly derived  in the 
noncanonical Hamiltonian representation \cite{Kuroda:1991}.
The application of Nambu mechanics
in meteorology was put forward
in recent years \cite{Nevir:2009, Sommer:2009}.

The continuous Rayleigh-B\'{e}nard convection
can be recast in a Nambu system  
using total energy and a Casimir that is 
considered resulting from Kelvin's circulation theorem
\cite{Bihlo:2008} or as 2D analogue of helicity 
\cite{Salazar:2010}.
Viscous dissipation is present
in a symmetric bracket using the difference between
total energy and unavailable potential energy.
Symmetric brackets have been proposed as
a natural extension of symplectic brackets
and the combined system has been coined 
'metriplectic' \cite{Morrison:1986}
(combining metric and symplectic systems).

Saltzman derived the Fourier mode truncations
of the 2D Rayleigh-B\'{e}nard convection in the Boussinesq approximation
\cite{Saltzman:1962}.
The three-mode truncation of these equations  
in the classical Lorenz-63 model \cite{Lorenz:1963} includes
a mode for the stream-function, a mode for temperature perturbation,
and a mode for the vertical stratification.
Lucarini and Fraedrich \cite{Lucarini:2009}
included heating by viscous dissipation
characterized by the Eckert number $Ec$
in a low-order model without shear flow.
In order to incorporate this 
effect, the Lorenz-63 model has to be extended to 
10 degrees of freedom including four stream function, 
four temperature perturbation, and two vertical 
stratification modes.

Some of the results presented in \cite{Lucarini:2009} 
suggest that the consideration of the Ec terms is relevant. 
First, it was found that when Ec=0 the system obeys a 
symmetry, which, by Noether's Theorem, corresponds to a 
conserved quantity. Such a symmetry is broken when 
positive values of Ec are considered, with the result 
that the phases of the temperature and streamfunctions 
are efficiently mixed and degenaracies are removed in 
the system, so that only one vanishing Lyapunov 
exponent is found. 
In \cite{Lucarini:2011_RG}, 
it was proposed that neglecting the heating 
effect due to viscous dissipation could be responsible 
for the observed imbalance of the global energy balance 
of the state-of-the-art global climate models. 

The dynamical system given by the Lorenz-63 equations
can be decomposed in a conservative part which conserves
volume in phase space and a divergent part 
related to forcing and dissipation.
The conservative part has two 
conserved quantities, a Hamiltonian $H$ and a Casimir $C$, 
which can be used to determine
the dynamics in a three-dimensional Nambu system
\cite{Nambu:1973, Nevir:1994} so that
$(\dot{X},\dot{Y},\dot{Z}) = \nabla C \times \nabla H + \nabla \Phi$,
where we have used the classical notation.
Forcing and dissipation is included in the gradient
of $\Phi$.
Thus Nambu dynamics provides
a clear geometric view of the nonlinear dynamics
and yields hints to the geometry of the strange
attractor in the forced and dissipative equations  
\cite{Nevir:1994, Axenides:2010}.
The Nambu structure of a six mode truncation of  
Saltzman's \cite{Saltzman:1962} convection model
has been presented recently \cite{Bihlo:2011}.
The derived model has 5 modes in common with the model studied here.

The aims of this publication are 
(i) 
to present the Nambu representation of the 
conservative part of the 10 degrees model of 
Rayleigh-B\'{e}nard convection presented in
\cite{Lucarini:2009}
and 
(ii) to determine the algebraic properties of 
heating by viscous dissipation. 
In Section \ref{TruSys} the truncated dynamical system
is presented. The conservation laws are derived in Section \ref{ConLaw}
with a physical interpretation in Section  \ref{PhyInt}.
The dynamics is represented in terms of a Nambu system using two conservation
laws  (Section \ref{NamStr}) from which a noncanonical Hamiltonian 
structure can be derived (Section \ref{HamStr}).
The viscous heating effect is shown to possess a symplectic 
as well as a metric structure (Section  \ref{VisHea}). 
Results are summarized and discussed in Section  \ref{SumCon}.

\section{Truncated equations}  \label{TruSys}

The physical setting is the two-dimensional Rayleigh-B\'{e}nard 
convection driven by a vertical temperature contrast 
$\Delta T$ across the height $H$. 
Dynamics in the vertical plane $(x,z)$
follows the Boussinesq approximation 
for incompressible velocity  
$u=-\partial_z \psi, v=\partial_x \psi$,
with the stream-function $\psi$,
and the linearized equation of state $\rho=\rho_0 (1-\alpha T)$
with the thermal expansion coefficient $\alpha$.
Temperature is decomposed 
according to 
$T=T_0 - z\Delta T/H + \theta$
in a mean temperature $T_0$,
a linear temperature gradient
and the perturbation $\theta$.
Boundary conditions at $z=0$ and $z=H$ along the $x$-axis
demand
vanishing vertical velocity, $\psi=0$,
vorticity, $\nabla^2 \psi=0$,
and temperature perturbation, $\theta=0$.


The main extension of the Boussinesq equations
suggested by \cite{Lucarini:2009} is the inclusion of
viscous heating in the temperature balance
    \begin{eqnarray} \label{RayBen}
		\partial_t \nabla^2 \psi 
		+ J(\psi,\nabla^2 \psi)
		= g \alpha \partial_x \theta
		+ \nu \nabla^4 \psi 
		\\
		\partial_t \theta + J(\psi,\theta)
		= \frac{\Delta T}{H} \partial_x \psi
		+ k \nabla^2 \theta
		+ \frac{\nu}{C_v}
		\partial_{ij} \psi  \partial_{ij} \psi
    \end{eqnarray}
here $\nu$ is viscosity, $k$ heat conductivity, 
$C_v$ is heat capacity,
and $g$ is gravity acceleration.
Non-dimensional equations are obtained by 
$x=Hx^\prime$, 
$t=H^2 t^\prime/k$, $\psi=k \psi^\prime$ 
and $\theta = k\nu \theta^\prime/g \alpha H^3$.
   
The impact of heating by viscous dissipation in the Fourier decomposition 
requires at least two modes for the
stream-function $\Psi_{n,m}$ and the temperature perturbation $\theta_{n,m}$,
furthermore, two modes for the perturbation of the stratification 
need to be included, $\theta_{0,2}$ and $\theta_{0,4}$ \cite{Lucarini:2009}.
The Fourier modes are associated with the wave vectors
($\pi a m, \pi n$).
Shear flow is excluded here by $\Psi_{0,1}=0$.
Real and imaginary parts of the stream-function
are split according to  
$\Psi_{1,1} = \alpha(X_1 + iX_2)$, $\Psi_{2,2} = \alpha(A_1+iA_2)$,
and the temperature perturbation is
$\theta_{1,1}=\beta (Y_1+iY_2)$, 
$\theta_{2,2}=\beta (B_1+iB_2)$, 
$\theta_{0,2} =i \gamma Z_1$, and 
$\theta_{0,4} =i \gamma Z_2$.

Heating by viscosity is proportional to the Eckert number,  
$Ec = k^2/(C_v \Delta T H^2)$ and leads to new nonlinear 
terms in the temperature equations of the Lorenz system. 
The present analysis concentrates on the conservative parts of the
dynamical equations with zero divergence in phase space.
The conservative parts are obtained by an operator splitting
through the neglect of the diagonal linear forcing terms
(see the general decomposition (4) in \cite{Lorenz:1963}). 

The conservative parts of the truncated equations 
retain all nonlinearities
    \begin{eqnarray} \label{dXYZdt}
		\dot{X}_1 &=& \sigma Y_2    		
													\label{dX1dt} \\
		\dot{X}_2 &=& -\sigma Y_1   		
													\label{dX2dt} \\
		\dot{Y}_1 &=& X_2 Z_1 - r X_2 
		+ 5/(\sqrt{2}\pi) \sigma Ec (X_1 A_1 + X_2 A_2) 
													\label{dY1dt} \\
		\dot{Y}_2 &=& -X_1 Z_1 + r X_1 
		+ 5/(\sqrt{2}\pi) \sigma Ec (X_1 A_2 - X_2 A_1) 
													\label{dY2dt} \\
		\dot{A}_1 &=& \sigma/2  B_2  		
													\label{dA1dt} \\
		\dot{A}_2 &=& -\sigma/2 B_1  		
													\label{dA2dt} \\
		\dot{B}_1 &=& 4A_2 Z_2 -2 r A_2 
		- 1/(2\sqrt{2}\pi) \sigma Ec (X_1^2 - X_2^2)  
													\label{dB1dt} \\
		\dot{B}_2 &=& -4A_1 Z_2 + 2r A_1 
		- 1/(\sqrt{2}\pi) \sigma Ec X_1 X_2     
													\label{dB2dt} \\
		\dot{Z}_1 &=& X_1 Y_2 - X_2 Y_1 
													\label{dZ1dt} \\
		\dot{Z}_2 &=& 4 A_1B_2 - 4 A_2 B_1 
													\label{dZ2dt}
    \end{eqnarray}    
The non-conservative terms not considered here
can be represented as a linear gradient, 
$-\nabla_x \Phi$, of the potential 
\begin{equation}
\Phi=\sigma (X_1^2 + X_2^2+4A_1^2+4A_2^2) 
 + Y_1^2 + Y_2^2
 + B_1^2 + B_2^2 
 + bZ_1^2+4bZ_2^2 
\end{equation} 
with respect to the 10 variables (\ref{dX1dt}) - (\ref{dZ2dt}), 
see \cite{Nevir:1994}. 
The notation follows the standard Lorenz-63 model
with the Prandtl number $\sigma=\nu/k$ 
and the relative Rayleigh number $r=R/R_c$, with
$R = g \alpha H^3 \Delta T/k \nu$, $R_c= 27\pi^4/4$,
and a geometric factor $b$; 
for further details see \cite{Saltzman:1962,Lucarini:2009}.
The symmetries and the instability of these equations 
have been intensively studied in \cite{Lucarini:2009}.

\section{Conservation laws}  \label{ConLaw}

Due to the absence of linear forcing and dissipative terms
in (\ref{dX1dt}) - (\ref{dZ2dt})
the phase space divergence vanishes and 
the equations satisfy the Liouville theorem. 
The equations have several time independent 
conservation laws (determined by the 
Reduce package CONLAW1 \cite{Wolf:2004}).

The first conservation law which is preserved 
for arbitrary Eckert number $Ec$ 
represents the sum of kinetic and potential energy and is denoted as 
Hamiltonian
    \begin{equation} \label{Hamiltonian}
		H = \frac{1}{2}
		\left(
		 X_1^2+X_2^2
		+4A_1^2+4A_2^2
		-2\sigma Z_1
		- \sigma Z_2
		\right)
    \end{equation}
Further conservation laws are found for $Ec=0$: The first is 
quadratic in the temperature perturbations 
    \begin{equation} \label{Casimir}
		C = \frac{1}{2}
		\left(
		 Y_1^2+Y_2^2
		+B_1^2+B_2^2
		+ Z_1^2 + Z_2^2
		-2rZ_1-rZ_2
		\right)
    \end{equation}
and the second is based on a coupling between stream-function and
temperature perturbation
    \begin{equation} \label{Casimir2}
		\hat{C} = \frac{1}{2}
		\left(X_1 Y_1 + X_2 Y_2 + A_1 B_1 + A_2 B_2
		\right)
    \end{equation}
The conserved quantities $C$ and $\hat{C}$ are denoted as Casimir
functions according to there specific interpretation in
Hamiltonian dynamics (see Section \ref{HamStr}).  

The Hamiltonians and the Casimirs for the two 
wave numbers considered are conserved individually
(pairs of variables $X,Y$ for $(n,m) = (1,1)$, and ($A,B$) for $(2,2)$).
The sum of $H$ and $C$ has been  analysed in
a related truncation \cite{Howard:1986,Thiffeault:1996} 
(denoted as "$Q$").  

The time derivative of the Casimir $C$ is linear in $Ec$
    \begin{eqnarray}  \label{dCdt} 
		\frac{dC}{dt} 
		&=& 
		\frac{\sigma Ec}{2 \sqrt{2} \pi}
		[
		10 Y_1 (A_1 X_1 + A_2 X_2)
		\nonumber
		\\
      &+&  10 Y_2 (A_2 X_1 - A_1 X_2)
		- B_1 (X_1^2 - X_2^2) - 2 B_2 X_1 X_2 
		]
    \end{eqnarray}
The time derivative of the second Casimir $\hat{C}$ is 
determined by the $Ec$-dependent dynamics of $B_1$ and $B_2$ only
    \begin{equation}  \label{dC2dt}
		\frac{d\hat{C}}{dt} 
		=
		\frac{9 \sigma Ec}{2 \sqrt{2} \pi}
		\left( A_1 ( X_1^2  - X_2^2) + 2 A_2 X_1 X_2
		\right)
    \end{equation}
The time derivative of $C$ in (\ref{dCdt}) will be used to formulate the 
$Ec$-dependent parts in terms of a gradient system 
(see the forthcoming Section \ref{GraSys}).

\section{Physical interpretation of the conservation laws} 
                                                  \label{PhyInt}

The total energy is the sum kinetic and  potential energy
    \begin{equation}  \label{Htot}
		H = KE + PE
    \end{equation}
which are given as integrals in the vertical plane
    \begin{equation}  \label{KEphys}
		KE  = \frac{1}{2} \int (\nabla \psi)^2 \mbox{d}V
    \end{equation}
    \begin{equation}  \label{PEphys}
		PE  = -\alpha g \int z \theta \mbox{d}V
    \end{equation}
(see for example \cite{Gluhovsky:1999}).

In the present truncation the kinetic energy $KE$ is given by 
    \begin{equation} \label{KE}
		KE = \frac{1}{2}
		\left(
		 X_1^2+X_2^2
		+4A_1^2+4A_2^2
		\right)
    \end{equation}
and total potential energy is
   \begin{equation} \label{PE}
	PE 
	 = -\frac{\sigma}{2}(2Z_1+Z_2)
	\end{equation}
The available potential energy characterizes the amount 
of potential energy which can be converted to kinetic energy
\cite{Lorenz:1955,Saltzman:1962}
    \begin{equation}  \label{APEphys}
		APE  = -  \frac{\alpha g H}{2 \Delta T} 
		\int \theta^2 \mbox{d}V
    \end{equation}
The truncated version is
\begin{equation} \label{APE}
		APE = -\frac{\sigma}{2r}
		\left(
		 Y_1^2+Y_2^2
		+B_1^2+B_2^2
		+ Z_1^2 + Z_2^2
		\right)
    \end{equation}
Note that $APE \leq 0$ since any perturbation reduces
the energy transformable to kinetic energy.
Total energy and the Casimir $C$ 
are related to $KE$ and $APE$ by
    \begin{equation}  \label{HKEAPE}
		H = KE + APE + \frac{\sigma}{r}C
    \end{equation}
Therefore, the Casimir contribution
can be identified as unavailable potential
energy \cite{Gluhovsky:1999}
   \begin{equation} \label{PEunav}
	\frac{\sigma}{r} C
	= PE - APE
	\end{equation}

The balance between kinetic and available potential energy is 
given by the Lorenz energy cycle 
    \begin{equation}  \label{dKEdt}
		\frac{d}{dt} KE 
		= \sigma
			\left(
				X_1 Y_2 - X_2 Y_1 + 2A_1 B_2 - 2 A_2 B_1
 			\right)
	 \end{equation}
    \begin{equation}  \label{dAPEdt}
		\frac{d}{dt} APE
		= - \frac{d}{dt} KE 
 			- \frac{\sigma}{r}  \frac{dC}{dt}
	 \end{equation}
For finite Eckert number ($Ec>0$)
viscous heating modifies available potential energy
(see $dC/dt$ in  (\ref{dCdt})). 
The balance between
$KE$ and $APE$ is closed, $dKE/dt + dAPE/dt=0$,
in the absence of viscous heating. 
The sum $E$ of kinetic and available potential energy
is the available energy
   \begin{equation}  \label{EKEAPE}
		E = KE + APE
    \end{equation}
and will be used represent the viscous heating term (see 
(\ref{dFCE}) below).

The Casimir functions (\ref{Casimir}, \ref{Casimir2}) 
correspond to the

Continuous versions of the Casimir functions 
are \cite{Bihlo:2008,Salazar:2010} 
$C_1 =  \int g(T-z) \mbox{d}V$ 
and $C_2 =  \int \zeta h(T-z) \mbox{d}V$, where
$T$ is the scaled deviation of the temperature from a linear 
profile $z$, $\zeta$ is vorticity; 
the integrals are in the vertical plane. 
$C_1$ is based on the  preservation of $T-z$ contours 
and corresponds to $C$ in (\ref{Casimir}) for the function $g(x)=x^2/2$.
$C_2$ depends on Kelvin's circulation theorem and 
corresponds to $\hat{C}$ in (\ref{Casimir2}).
In the continuos Nambu representation of \cite{Bihlo:2008} 
the Casimir $C_2$ is used together with the Hamiltonian
$H = \int (1/2 (\nabla \psi)^2-Tz)\mbox{d}V$, 
while \cite{Salazar:2010} compare different Nambu
representations using $H$, $C_1$ (for $h(x)=x$), enstrophy and  
total buoyancy ($C_1$ with $g(x)=x$).
In the present analysis the first Casimir 
$C$ in (\ref{Casimir}), i.e. 
the unavailable potential energy is used in order to continue previous work
on the Nambu representation of the Lorenz-63 model \cite{Nevir:1994}.
A further motivation for using this Casimir
is its role in the viscous heating (see Section \ref{GraSys}).

\section{Nambu structure}  \label{NamStr}

N\'{e}vir and Blender \cite{Nevir:1994} developed a Nambu 
representation of the 
non-dissipative parts of the classical Lorenz-63 \cite{Lorenz:1963} model
for $X_1,Y_2$, and $Z_1$ 
(see also \cite{Axenides:2010, Pelino:2010, Bihlo:2011} and \cite{Roupas:2011} 
for a classification). 
With the conservation laws for $H$ and $C$ 
the Lorenz equations are in terms of $x = (X_1,Y_2,Z_1)$
    \begin{equation}  \label{Nambu3}
		\frac{d}{dt} x_i
		= - \varepsilon_{ijk}
		C_j H_k,
		\qquad i,j,k = 1,2,3
    \end{equation}
with the Levi-Civita symbol $\varepsilon_{ijk}$
(note that
the definitions in \cite{Nevir:1994} are interchanged). 
The second Casimir $\hat{C}$ (\ref{Casimir2}) 
does not occur in the classical Lorenz-63
truncation since $Y_1$ and $X_2$ are not considered.

Rewriting the dynamical system (\ref{dX1dt}) - (\ref{dZ2dt})
for the 10-component variable vector 
    \begin{equation} \label{x}
		x = (X_1,Y_2,Z_1,X_2,Y_1,A_1,B_2,Z_2,A_2,B_1)
    \end{equation}
the Nambu equations for the extended Lorenz system is 
    \begin{equation}  \label{Nambu10}
		\frac{dx}{dt} 
		= {\mathcal N} (\nabla_x C, \nabla_x  H) 
    \end{equation}
with the Nambu tensor ${\mathcal N}$ and the gradient $\nabla_x$.

The non-vanishing elements of the Nambu tensor are
    \begin{equation}  \label{Nambuijk}
		N_{1,2,3} = -\varepsilon_{1 2 3}, \quad 
		N_{3,4,5} =  \varepsilon_{3 4 5}, \quad
		N_{6,7,8} = -\varepsilon_{6 7 8}, \quad 
		N_{8,9,10} = \varepsilon_{8 9 10}
    \end{equation}
with the standard definition of the Levi Civita symbol $\varepsilon$.
Thus the Nambu tensor is cyclic and
anti-symmetric.
Furthermore, it satisfies the generalized
Jacobi identity \cite{Takhtajan:1994,Sommer:2011}
    \begin{equation}  \label{GenJacIde}
		N_{ijk} N_{lpq} +
		N_{ijq} N_{lkp} +
		N_{ijp} N_{lqk} = 0
    \end{equation}
in the form required for constant Nambu tensors.


Based on (\ref{Nambu10}) a Nambu bracket for arbitrary 
phase space functions $F(x)$ is written
   \begin{equation} \label{Nambucyc}
		\frac{dF}{dt}
		= \{F,C,H \}
	\end{equation}
which is cyclic $\{F,C,H \}=\{C,H,F \}=\{H,F,C \}$.
This cyclicity is the main ingredient in applications
of the Nambu representation in numerical models \cite{Salmon:2005}.
This bracket vanishes for the second Casimir (\ref{Casimir2}), 
$\{\hat{C},C,H \}=0$.

The Nambu representation of the conservative terms
can be presented in terms of brackets 
    \begin{equation}  \label{Nbar}
		\frac{d}{dt}
       \left(
          \begin{array}{c}
          	X_1 \\
          	Y_2 \\
          	Z_1 \\
          	X_2 \\
          	Y_1 \\
          	A_1 \\
          	B_2 \\
          	Z_2 \\ 
          	A_2 \\
          	B_1 \\
           \end{array}
       \right)
		=
       \left(
          \begin{array}{rrrr}
            \{Z_1,Y_2\} &             &            &             \\
            \{X_1,Z_1\} &             &            &             \\
            \{Y_2,X_1\} & +\{X_2,Y_1\}  &            &             \\
                       &   \{Y_1,Z_1\}  &            &             \\
                       &   \{Z_1,X_2\}  &            &             \\
                       &             & \{Z_2,B_2\}  &             \\
                       &             & \{A_1,Z_2\}  &             \\
                       &             & \{B_2,A_1\}  & +\{A_2,B_1\}  \\
                       &             &            &  \{B_1,Z_2\}  \\
                       &             &            &  \{Z_2,A_2\}  \\
           \end{array}
       \right)
    \end{equation}
where the bracket $\{,\}$ is the Jacobian of $C$ and $H$
with respect to $a$ and $b$, 
$\{a,b\} = \partial C/\partial a - \partial H/\partial b $.
The three brackets in the upper left 
represent the Nambu dynamics of the Lorenz-63 (\ref{Nambu3})
model \cite{Nevir:1994}.
This structure demonstrates the decoupling of the two wave number
modes $(m,n) = (1,1)$ and $(2,2)$ in the five upper 
and the five lower rows.
Furthermore, the coupled Fourier modes are separated in
the columns.
Coupling between the wave number modes is given by the 
feedback represented in the Eckert terms which is not included here
(see Section \ref{VisHea}).

Instead of the Hamiltonian the dynamical equations 
can be rewritten using the available energy 
$E=KE+APE$  
in the Nambu bracket  (\ref{Nambucyc})
   \begin{equation}  \label{dFCE}
		\frac{dF}{dt}
		= \{F,C,E \} 
	\end{equation}
since $E$ deviates from $H$
only by the Casimir 
    \begin{equation}
		H = E + \frac{\sigma}{r}C
    \end{equation}
which is annihilated in the bracket.

\section{Hamiltonian structure} \label{HamStr}

The Hamiltonian structure of the dynamical system 
(\ref{dX1dt}) - (\ref{dZ2dt})
is a
straightforward evaluation of the Nambu system (\ref{Nambu10})
(also for $Ec=0$)
    \begin{equation}  \label{dxdtP}
		\frac{d}{dt}x
		= {\mathcal P} \, \nabla_x H
		= {\mathcal N} (\nabla_x C, \nabla_x H) 
    \end{equation}
with the anti-symmetric Poisson tensor
    \begin{equation}  \label{PNDC}
      {\mathcal P} (.)
		= {\mathcal N} (\nabla_x C, .) 
    \end{equation}
The elements of the Poisson tensor are 
explicitly given by derivatives of the Casimir
with respect to the variables in (\ref{x}),
for example $C_1 = \partial C/\partial X_1$, etc.
    \begin{equation}  \label{P}
       {\mathcal P} =
       \left(
          \begin{array}{cccccccccc}
              0 &  C_3 & -C_2 &    0 &    0 & 0 & 0 & 0 & 0 & 0   \\
           -C_3 &    0 &  0   &    0 &    0 & 0 & 0 & 0 & 0 & 0   \\
            C_2 &    0 &    0 & -C_5 &    0 & 0 & 0 & 0 & 0 & 0   \\
              0 &    0 &  C_5 &    0 & -C_3 & 0 & 0 & 0 & 0 & 0   \\
              0 &    0 &    0 &  C_3 &    0 & 0 & 0 & 0 & 0 & 0   \\
              0 &    0 &    0 &    0 &    0 & 0 & C_8 & -C_7 & 0 & 0   \\
              0 &    0 &    0 &    0 &    0 & -C_8 & 0 & 0 & 0 & 0   \\
              0 &    0 &    0 &    0 &    0 & C_7 & 0 & 0 & -C_{10} & 0 \\
              0 &    0 &    0 &    0 &    0 & 0 & 0 & C_{10} & 0 & -C_8 \\
              0 &    0 &    0 &    0 &    0 & 0 & 0 &  0 & C_8 & 0   \\
           \end{array}
       \right)
    \end{equation}

The Poisson tensor ${\mathcal P}$ defines a 
Poisson bracket for the dynamics of arbitrary 
phase space functions $F(x)$ 
   \begin{equation}  \label{dFdtFH}
		\frac{dF}{dt}
		= \{F,H \}_P
	\end{equation}
The Casimir $C$ is conserved as the nullspace 
of the Poisson tensor
    \begin{equation}  \label{PDC}
	 	{\mathcal P} \, \nabla_x C  = 0
    \end{equation}
Thus, in Hamiltonian dynamics the conservation of Casimirs
is an inherent property hidden in the Poisson tensor.
These conservation laws lead to foliations in phase space
and are interpreted as a kinematic property of the dynamical system,
whereas the conservation of the Hamiltonian is considered as
a dynamic property.

\section{Viscous heating by Eckert terms}   \label{VisHea}

An important property of the $Ec$-dependent terms in the equations 
for $Y_1, Y_2, B_1$, and $B_2$ (\ref{dY1dt}), (\ref{dY2dt}),
(\ref{dB1dt}), and (\ref{dB2dt})
is
that they satisfy the Liouville theorem.
Thus a representations in terms of a stream-function 
is possible (hence as Hamiltonian systems). 
To obtain this we consider the equations 
for $Y_1, Y_2$ and $B_1, B_2$ in two complex plains
given by $Y=Y_1+iY_2$ and $B=B_1+iB_2$. 
The dynamic equations are 
determined by the gradient of the complex function 
   \begin{equation}  \label{fYB}
	 	f(Y,B) =
	 	\frac{\sigma Ec}{2 \sqrt{2} \pi} 
	 	\left( 10 X^* A Y^*
	 	-
	 	X X B^*
	 	\right)
    \end{equation}
with $X=X_1+iX_2$ and $A=A_1+iA_2$.
The Cauchy-Riemann equations yield the dynamics for $Y$ 
and $B$ in terms of a stream-function (Im $f$) 
and the gradient of potential (Re $f$).
It turns out that the potential is given by the
time derivative of the Casimir $C$ (unavailable potential energy)
which leads to a metric system \cite{Morrison:1986}.
The stream-function is considered as a second Hamiltonian
$\tilde{H}$ (not to be confused with (\ref{Hamiltonian})),
thus the complex function is $f=\dot{C}+i\tilde{H}$.
In the following the Hamiltonian and the gradient 
representation are considered.

\subsection{Hamiltonian system}  \label{HamSys}

The Hamiltonian $\tilde{H}=$ Im$f$ (\ref{fYB})
can be constructed by the 
Ec-dependent forcing terms for the set of 
variables $y_j=\{Y_2, Y_1, B_2, B_1\}$ in (\ref{x})
  \begin{equation} \label{dyG}
   \frac{d}{dt} y_j  
   =G[y_j]
	\end{equation}
using that these terms $G$ depend only on $X_1, X_2, A_1, A_2$.
Therefore, $H$ is formally
   \begin{equation} \label{Psi}
       \tilde{H} = Y_2 G[Y_1] - Y_1 G[Y_2] + B_2 G[B_1] - B_1 G[B_2]
   \end{equation}
which can be rearranged in the compact representation 
   \begin{equation} \label{Psify}
       \tilde{H} = Y_2^2 \frac{d}{dt} 
                    \frac{Y_1}{Y_2} 
              +
              B_2^2 \frac{d}{dt} 
                    \frac{B_1}{B_2} 
   \end{equation}
determined by the phases of 
$\theta_{1,1} = \beta (Y_1+i Y_2)$ and 
$\theta_{2,2} = \beta (B_1 + iB_2)$.
In terms of the dynamical variables, $\tilde{H}$ has the 
explicit form
   \begin{eqnarray} \label{Psivars}
      \tilde{H} 
      &=& \frac{\sigma Ec}{2 \sqrt{2} \pi}
      [
      10 Y_2 (A_1 X_1 + A_2 X_2) 
      \nonumber
      \\
      &-& 10 Y_1 (A_2 X_1 - A_1 X_2)
      - B_2( X_1^2 - X_2^2) + 2 B_1 X_1 X_2
      ]
	\end{eqnarray}
Thus the viscous heating effect can be represented
by a second Hamiltonian part
   \begin{equation} \label{dyPEc}
   \frac{d}{dt} y   
   = {\mathcal P}_{Ec} \nabla_y \tilde{H}
	\end{equation}
where $\nabla_y$ is the gradient with respect to $y$. 
The Poisson tensor is canonical
with the non-zero elements
	\begin{eqnarray}
		{\mathcal P}_{Ec}(Y_1,Y_2) & =& -1, \quad
		{\mathcal P}_{Ec}(Y_2,Y_1) = 1 
		\nonumber
		\\
		{\mathcal P}_{Ec}(B_1,B_2) & =& -1, \quad
		{\mathcal P}_{Ec}(B_2,B_1) = 1
	\end{eqnarray}
%

The dynamics (\ref{dyPEc}) has a canonical Poisson bracket 
   \begin{equation} \label{Pecbra}
		\{ F,\tilde{H} \}_{Ec} 
		= \nabla_x F {\mathcal P}_{Ec} \nabla_x \tilde{H}
	\end{equation}
which conserves the Hamiltonian (\ref{Hamiltonian}), 
$ \{ H,\tilde{H} \}_{Ec} = 0 $.
Thus $H$  can formally be considered as a Casimir 
of the tensor ${\mathcal P}_{Ec}$,  ${\mathcal P}_{Ec} H =0$.

\subsection{Gradient System}   \label{GraSys}

The $Ec$-dependent terms in (\ref{dX1dt}) - (\ref{dZ2dt})
are the gradient of a potential given by the
real part Re$f$  (\ref{fYB}).
This potential can be identified as 
time derivative of the Casimir (\ref{dCdt}).
Hence the present low order system 
shows a relationship between the
Hamiltonian part (\ref{P}) and the viscous heating
mediated by the Casimir (\ref{Casimir}).
The potential can be constructed from the Ec-dependent
forcing terms (\ref{dyG}) according to 
   \begin{equation} \label{CYf}
       \frac{dC}{dt} = Y_1 G[Y_1] + Y_2 G[Y_2] + B_1 G[B_1] + B_2 G[B_2]
   \end{equation}
Embedding this in the dynamics of $x$ the gradient 
representation is
    \begin{equation}  \label{dxdtM}
		\frac{d}{dt}x
   	 = {\mathcal M} \nabla_x \frac{dC}{dt}
    \end{equation}
with the symmetric operator ${\mathcal M}=diag(0,1,0,0,1,0,1,0,0,1)$
which selects the variables $Y_2, Y_1, B_1, B_2$ in the complete 
10-component vector $x$ (\ref{x}).
Thus for an arbitrary phase space function $F$ the 
contribution of viscous heating 
(\ref{dxdtM})
can be included in a symmetric bracket (also denoted as metric bracket
in contrast to symplectic Poisson brackets).
   \begin{equation} \label{FWbra}
		\langle F, \frac{dC}{dt} \rangle
		= \nabla_x F {\mathcal M} \nabla_x \frac{dC}{dt}
	\end{equation}
Due to the Cauchy-Riemmann equations the $\tilde{H}$ manifolds are
perpendicular to $\dot{C}$ and $\tilde{H}$ is preserved in this bracket, 
$\langle \tilde{H}, \dot{C} \rangle =0$.
With (\ref{dFCE}) the dynamics of the full
system can be completely understood in terms of the available energy $E$
\begin{equation}   \label{dFCEEc}
		\frac{dF}{dt}
		= \{F,C,E \} 
		- \frac{r}{\sigma} \langle F,\frac{dE}{dt} \rangle
	\end{equation}
since  
	\begin{equation}   \label{WdEdt}
		\frac{dC}{dt} 	= - \frac{r}{\sigma} \frac{dE}{dt}
	\end{equation}
represents the change in available energy.

This symmetric bracket does not represent physical dissipation 
associated with the convergence of trajectories in phase space. 
The reason is that the dissipated energy is re-introduced 
in the temperature equation which is represented
by dynamical degrees of freedom.

A main property of this symmetric bracket is that it
generates a positive change of the Casimir tendency 
   \begin{equation}   \label{ddCdtdt}
		\langle \dot{C},  \dot{C} \rangle
		= \frac{\sigma^2 Ec^2}{8 \pi^2}
		\left( 100(A_1^2+A_2^2)(X_1^2+x_2^2) + (X_1^2+X_2^2)^2 \right)
	   \geq 0
	\end{equation}
note that this is proportional to $Ec^2$. 
Since the Nambu bracket representing the conservative part
does not vanish  $ \{\dot{C},C,E \}  \neq 0 $,
the time derivative of the Casimir tendency 
in the full equations has no definite sign
   \begin{equation}   \label{ddCdtdtCE}
		\frac{d}{dt}
		\dot{C}
		= \{\dot{C},C,E \}  + \langle \dot{C},  \dot{C} \rangle
	\end{equation}
Due to this property, $\dot{C}$ cannot be 
considered as the analogue to entropy
(see the concept in \cite{Kaufman:1984}).


Since the Hamiltonian is conserved for arbitrary Eckert number
the symmetric bracket vanishes for $H$
   \begin{equation}   \label{dHdt}
		\frac{dH}{dt}
		= \langle H,  \dot{C} \rangle
		= 0
	\end{equation}
Hence the Hamiltonian can be considered as a 
nullvector of the symmetric bracket \cite{Kaufman:1984} 
(see the construction method \cite{Morrison:1986}).

The role of the Casimir is twofold: 
(i) As a conserved quantity it determines the 
non-viscous dynamics in the Nambu representation, and 
(ii) the time derivative determines the heating effect 
in a gradient system.

\section{Summary and Conclusions}   \label{SumCon}

This paper discusses the impact of viscous heating on the
dynamic properties of convection.
The model is given by a truncation of the Boussinesq 
equations \cite{Saltzman:1962} with an additional
heating caused by viscous dissipation \cite{Lucarini:2009}. 
Wave numbers selected are $n=m=1$ and $n=m=2$ for the stream function
and the temperature perturbation, where additional vertical 
wave numbers $m=2$ and $m=4$ are included.
Vertical shear is not incorporated.
Linear forcing and friction are removed in the dynamical equations 
to obtain a divergence free dynamical system satisfying Liouville's theorem.

The truncated systen shows one conservation law for all $Ec$ 
which is given by the sum of kinetic and potential energy
and identified as the Hamiltonian of the dynamical system.
There are two quadratic conservation laws for vanishing 
Eckert number, $Ec=0$, a first, $C$, related to
unavailable potential energy (UPE, the residuum between total energy and
available potential energy) and a second, $\hat{C}$,
related to the coupling of stream-function and temperature
perturbations.
Both conservation laws are Casimirs of the Hamiltonian representation.
The viscous heating changes both Casimir functions.
An available potential energy APE is identified in the truncated
system which is convertible to kinetic energy.
The Lorenz energy cycle for kinetic and available
potential energy has a source term linear in $Ec$.

A Nambu representation is defined using the Hamiltonian
$H$ and the Casimir $C$ to determine the time derivative 
of an arbitrary state space function by a Nambu bracket,
$ \dot{F} = \{F, C, H \}$.
The Nambu bracket is cyclic and 
satisfies a generalized Jacobi identity.
In a Nambu representation the Casimirs can be considered
as additional Hamiltonians. 
The Nambu representation can be directly 
transformed into noncanonical Hamiltonian dynamics 
by evaluating the Casimir dependency in the Nambu bracket,
$ \{F, C, H \} = \{F, H \}_P$.


Casimirs and the Hamiltonian are excellent observables 
for the study of the response  of dynamical system to 
small perturbations far from equilibrium
\cite{Ruelle:1998,Lucarini:2008_JSP,Lucarini:2009_JSP}.
Even when these quantities are not conserved in the
full nonlinear equations, they provide an intrinsic skeleton 
characterizing the geometry of the phase space trajectories.
Promising results have been obtained recently for 
perturbations of the Lorenz 96 model considering 
total energy and momentum analogues \cite{Lucarini:2011_NPG}.

The final result of the present publication is
that the dynamics (\ref{dX1dt}) - (\ref{dZ2dt})
can be decomposed in a Nambu representation
and a metric representation of the 
viscous heating which is proportional to $Ec$
	\begin{equation}
 		\dot{F} = \{F, C, H \} 
		+ \langle F, \dot{C} \rangle
	\end{equation}
The Casimir $C$ appears in the Nambu as well as in the 
metric term (via the time derivative).
Furthermore, the viscous heating bracket 
is equivalent to a canonical Hamiltonian bracket,
$ \langle F, \dot{C} \rangle = \{F, \tilde{H} \}_{Ec}$,
due to the Cauchy-Riemann equations.

The present low order model revels specific
properties of the viscous heating which is usually neglected
in models:
\begin{enumerate}

\item 
Viscous heating does not perturb 
conservation of total energy (the Hamiltonian).
This result may have implications for considering this effect 
in complex models.

\item 
The viscous heating terms are divergence 
free in phase space, hence the full dynamical 
system satisfies the Liouville theorem.

\item 
The Lorenz energy cycle for kinetic
and available energy has a source term given
by viscous heating. This effect alters convective processes
by modifying available potential energy.

\item
Viscous heating determines the
time derivative of the Casimir in the Nambu representation.
Nevertheless, its role in the Nambu and the Hamiltonian representation 
is unaltered.

\item 
The viscous heating terms can be represented
by a symmetric bracket with a positive
change of the Casimir time derivative, 
$ \langle \dot{C}, \dot{C} \rangle \geq 0$. 
However, the interpretation of $\dot{C}$ as an entropy 
according to \cite{Kaufman:1984}
is not possible since the Poisson bracket (\ref{dFdtFH}) 
does not vanish, $\{\dot{C},H\}_P \neq 0$.

\end{enumerate}

This study shows that a correct energy recycling budget 
leads to conservation of total energy and retains
the Hamiltonian and Nambu structures.
The single impact is a change of the Casimir function,
without altering it's role in the Nambu representation
(such conservation laws have been termed constitutive 
in Nambu dynamics \cite{Nevir:2009,Salazar:2010}).
The role of the Casimir in the recycling process 
might be a particularity of the present model,
but the feedback in the Lorenz energy cycle
and the recreation of kinetic energy 
could shed new light on the relationships
with entropy \cite{Kaufman:1984, Morrison:1986}. 
The Ec-term mixes available potential energy (APE) 
and unavailable potential energy.
This implies that an improper treatment of the Ec-term can have a 
large impact on the energy cycle \cite{Lorenz:1955,Lucarini:2011_RG}
and bias considerably the thermodynamic efficiency of a geophysical
fluid
\cite{Lucarini:2011_RG,Johnson:1989,Johnson:2000,Lucarini:2009_PRE}.

\bibliographystyle{plain}

\bibliography{Nambu}

\end{document}